
\input phyzzx.tex
\input phyzzx.fonts
\tolerance=5000
\sequentialequations
\overfullrule=0pt
\twelvepoint
\nopubblock
\PHYSREV
\doublespace

\def\eps{\epsilon}
\def\geff{\Gamma_{\rm eff}}
\def\DD{{\cal D}}
\def\LL{{\cal L}}
\def\sla{\raise.16ex\hbox{$/$}\kern-.57em}
\def\Dsl{\kern.2em\raise.16ex\hbox{$/$}\kern-.77em\hbox{$D$}}
\def\parsl{\sla\hbox{$\partial$}}
\def\Asl{\kern.2em\raise.16ex\hbox{$/$}\kern-.77em\hbox{$A$}}
\def\Bsl{\kern.2em\raise.16ex\hbox{$/$}\kern-.77em\hbox{$B$}}
\def\square{\kern1pt\vbox{\hrule height 1.2pt\hbox{\vrule width 1.2pt\hskip 3pt
   \vbox{\vskip 6pt}\hskip 3pt\vrule width 0.6pt}\hrule height 0.6pt}\kern1pt}
\def\BF{B\wedge F}
\def\mn{{\mu\nu}}
\def\rs{{\rho\sigma}}
\def\mnrs{{\mn\rs}}
\def\bmn{B_\mn}
\def\bMN{B^\mn}
\def\hmnr{H_{\mn\rho}}
\def\hMNR{H^{\mn\rho}}
\def\fmn{F_\mn}
\def\fMN{F^\mn}
\def\am{A_\mu}
\def\an{A_\nu}

\def\ie{{\it i.e.},\ }

\gdef\journal#1, #2, #3, 1#4#5#6{               
    {\sl #1~}{\bf #2}, #3 (1#4#5#6)}            
\def\mpl{\journal Mod. Phys. Lett., }
\def\np{\journal Nucl. Phys., }
\def\pl{\journal Phys. Lett., }
\def\prl{\journal Phys. Rev. Lett., }
\def\pr{\journal Phys. Rev., }
\def\prd{\journal Phys. Rev. D, }

\def\ijmp{\journal Int. Jour. Mod. Phys., }

\line{\hfill UdeM-LPN-TH-162}
\line{\hfill CRM-1916}
%
\titlepage
\title{Induced $B\wedge F$ term and photon mass generation in 3+1
dimensions}
\author{M. Leblanc,$^{1,2}$ R. MacKenzie,$^1$ P.K.
Panigrahi,$^1${}\footnote{\hbox{$^*$}}{Address after 1 October, 1993:
Department of Physics, University of Hyderabad, Hyderabad, India, 500134.}
R. Ray$^{3,4}$}

\vskip.2cm\singlespace
\centerline{$^1${\sl Laboratoire de Physique Nucl\'eaire,
Universit\'e de Montr\'eal}}
\centerline{{\sl Montr\'eal, Qu\'e, Canada H3C 3J7}}
\smallskip
\centerline{$^2$ {\sl Centre de Recherches Math\'ematiques,
Universit\'e de Montr\'eal}}
\centerline{{\sl Montr\'eal, Qu\'e, Canada H3C 3J7}}
\smallskip
\centerline{$^3$ {\sl Physics Dept.,
City College of the City University of New York}}
\centerline{{\sl New York, NY 10013}}
\smallskip
\centerline{$^4$ {\sl Department of Physics, University of Maryland}}
\centerline{\sl College Park, MD 20742}
\doublespace
\abstract{
Analysing a 3+1 dimensional model with four-Fermi interactions, we show
that topological $\BF$ terms (both abelian and non-abelian) can be induced
radiatively by massive fermions at the one-loop level. It is further
pointed out that a mechanism of photon (or non-abelian gauge field)
mass generation distinct from the usual Higgs mechanism, through the $\BF$
term, is also implemented in the long-distance effective action of this
model, provided a gap equation is satisfied.
}

\endpage

\REF\kalram{M. Kalb and P. Ramond, \prd 9, 2273, 1974.}

\REF\deser{S. Deser, \pr 187, 1931, 1969.}

\REF\chern{R. Jackiw and S. Templeton, \prd 23, 2291, 1981; J. Schonfield,
\np B185, 157, 1981; S. Deser, R. Jackiw and S. Templeton, \journal Ann.
Phys., 140, 372, 1982.}

\REF\horowitz{G.T. Horowitz, \journal Comm. Math. Phys., 125, 417, 1989.}


\REF\mass{E. Cremmer and J. Scherk, \np B72, 117, 1974;
D.Z. Freedman and P.K. Townsend, \np B177, 282, 1981;
T. R. Govindarajan, \journal J. Phys. G. (Nucl. Phys.), 8, L17, 1982;
J.A. Minahan and R.C. Warner, Florida preprint UFIFT-HEP-89-15 (1989);
T.J. Allen, M. Bowick and A. Lahiri, \mpl A6, 559, 1991.}

\REF\namass{A. Lahiri, preprint LA-UT-92-3477 (1992).}

\REF\polyakov{A.M. Polyakov, \mpl A3, 325, 1988.}

\REF\harliu{X. Fustero, R. Gambini and A. Trias, \prl 62, 1964, 1989;
R. Gambini and L. Setaro, \prl 65, 2623, 1990;
J.A. Harvey and J. Liu, \pl 240B, 369, 1990;
A.P. Balachandran, V.P. Nair, B.-S. Skagerstam and A. Stern, \prd 26, 1443,
1982;
C. Aneziris, A.P. Balachandran, L. Kauffman and A.M. Srivastava, \ijmp A6,
2519, 1991.}

\REF\rajeev{S.G. Rajeev, preprint MIT-CTP-1335 (unpublished).}

\REF\balteo{A.P. Balachandran and P. Teotonio-Sobrinho, \ijmp A8, 723,
1993.}

\REF\greschwit{M.B. Green, J.M. Schwartz and E. Witten, {\sl Superstring
Theory}, Cambridge University Press, New York, 1987.}

\REF\redlich{A.N. Redlich, \prd 29, 2366, 1984.}

\REF\hs{J. Hubbard, \prl 3, 77, 1959;
R.L. Stratonovich, \journal Dokl. Akad. Nauk. SSSR, 115, 1097, 1957.}

\REF\odell{T.H. O'Dell, {\sl The Electrodynamics of magneto-electric
media}, North-Holland Publishing Company, Amsterdam, 1970.}

\REF\wenzee{X.G. Wen and A. Zee, preprint NSF-ITP-88-115 (1988).}

\REF\zee{A. Zee, preprint NSF-ITP-90-55 (1990).}

\REF\schakel{See
A.M.J. Schakel, {\sl On broken symmetries in Fermi systems}, U.
Amsterdam thesis (1989), and references therein; V.P. Nair, private
communication.}

\REF\aitfra{I.J.R. Aitchison and C.M. Fraser, \pl 146B, 63, 1984.}

In four dimensional gauge theories, one can contemplate adding a so-called
$\theta$-term proportional to $\eps^\mnrs \fmn F_\rs$
to the Lagrangian. At first sight, such a term appears
unimportant, being a total derivative and therefore having no effect on the
equations of motion, which are derived from local considerations.
However it has an influence on global aspects of the theory,
since it measures the topology of the gauge field
configuration. This gives rise to a variety of well-known effects such as
$\theta$-vacua and
CP-violation.

A similar term which does have a local (as well as global) significance can
be written in a theory where one has a vector and antisymmetric tensor
field. Consider the following
Lagrangian describing a U(1) gauge field $\am$ and
an antisymmetric tensor field $\bmn$ [\kalram]:
$$
\LL^{\rm free}=-{1\over4}\fmn\fMN+{1\over12}\hmnr\hMNR.
\eqn\un
$$
Here, $\fmn=\partial_\mu\an-\partial_\nu\am$ and
$\hmnr=\partial_\mu B_{\nu\rho}+\partial_\nu B_{\rho\mu}+
\partial_\rho B_{\mu\nu}$ are the respective field strengths for the gauge
and antisymmetric tensor fields. The Lagrangian is invariant under the
gauge invariances $\am\to\am+\partial_\mu\Lambda$ and
$\bmn\to\bmn+\partial_\mu\Lambda_\nu-\partial_\nu\Lambda_\mu$, where
$\Lambda$ and $\Lambda_\mu$ are scalar and four-vector gauge parameters,
respectively. A term which respects these two symmetries (up to a total
derivative) and which
therefore can be added to the Lagrangian above is the following
[\kalram,\deser], which we
refer to as the $\BF$-term:
$$
g\eps^\mnrs\bmn F_\rs.
\eqn\deux
$$
This term is a natural generalization of the topological Chern-Simons (CS)
term [\chern] to 3+1 dimensions, and as such has attracted considerable
interest [\horowitz]
in the literature recently for its potential to
encode topological information about four-manifolds, as the CS term has
done recently for three-manifolds. Furthermore, it has been shown recently
that the combined action \un+\deux\ provides a
mechanism for photon mass generation which is topological in nature and is
quite distinct from the usual Anderson-Higgs mechanism [\mass,\namass].
{}From another point of view, there have been attempts to generalize
Polyakov's construction [\polyakov] of the transmutation of the statistics of
point particles in 2+1 dimensions via a CS photon to that of strings in 3+1
dimensions in the presence of the $\BF$ term. Broadly speaking, the $\BF$
term can keep track of the way in which a string world sheet ``braids''
around the vortex world line, via a generalization of the Wilson loop:
$$
\langle e^{i\oint\am dx^\mu} e^{i\oint\bmn dS^{\mn}}\rangle
\eqn\trois
$$
Composites of one-dimensional objects and vortices can exhibit fractional
statistics and angular momentum in 3+1 dimensions [\harliu].

An antisymmetric tensor field appears naturally in the dual formulation
of free, massless scalar field theory [\kalram,\rajeev].
That this is true can be seen from the
following on-shell derivation. We begin with the massless Klein-Gordon
equation:
$$
\square\phi=0
\eqn\dog
$$
If we define the vector $R_\mu=\partial_\mu\phi$, \dog\ becomes
$$
\partial_\mu R^\mu=0.
\eqn\cat
$$
Thus, $R$ is divergence-free; \ie it can be expressed as a curl:
$$
R_\mu=\eps_\mnrs\partial^\nu B^\rs,
\eqn\fly
$$
with $B^\rs$ an antisymmetric tensor field. On the other hand, $R$ is the
gradient of $\phi$, so
$$
\partial_\mu R_\nu-\partial_\nu R_\mu=0.
\eqn\ant
$$
In terms of $B$, this is
$$
\partial_\mu\left(\partial^{[\mu}B^{\nu\rho]}\right)=0,
\eqn\cinq
$$
an equation which is invariant under the gauge transformation
$B_\mn\to\partial_\mu\Lambda_\nu-\partial_\nu\Lambda_\mu$ and which can be
derived from the Lagrangian
$$
\LL={1\over12}\left(\partial_\mu B_{\nu\rho}
+\partial_\nu B_{\rho\mu}+\partial_\rho
B_\mn\right)^2={1\over12}(H_{\mn\rho})^2.
\eqn\hey
$$

Another situation where the $\BF$
term naturally appears is in the dual formulation of the London
limit of the Landau-Ginzburg action [\balteo],
and also in string theories,
where they are instrumental in
implementing the anomaly cancellation mechanism of Green and Schwartz
[\harliu,\greschwit].
It is therefore of considerable importance to investigate
whether or not these terms could arise by other means.

In this paper, we show that such terms do indeed necessarily appear
at the one-loop level in some
familiar theories with four-fermi couplings. In this respect, it resembles
the radiative appearance of the CS term in the presence of a
parity-violating massive fermion
in 2+1 dimensions [\redlich]. As occurs there,
the $\BF$ term is the lowest order term in a derivative
expansion of the effective action, and it should play a dominant role in
determining the various low-energy phases of the theory.

The model studied is conventional massive electrodynamics with added
four-fermi terms of Thirring (current-current)
and dipole-dipole type. We begin with the
Lagrangian
$$
\LL=-{1\over4e^2}\fmn^2+\bar\psi (i\Dsl-m)\psi
-{g_1\over2}(\bar\psi\gamma^\mu\psi)(\bar\psi\gamma_\mu\psi)
-{g_2\over2}(\bar\psi\sigma^\mn\gamma_5\psi)(\bar\psi\sigma_\mn\gamma_5\psi),
\eqn\six
$$
with $D_\mu=\partial_\mu+i  A_\mu$.

The fermion functional integral is quite complicated due to the four-fermi
terms. However, it can be simplified using a standard trick known as a
Hubbard-Stratonovich transformation [\hs]. For each four-fermi term, one
introduces an auxiliary field: a vector field $a_\mu$ in the case of the
Thirring term, an antisymmetric tensor field $\bmn$ in the case of the
dipole-dipole term. Thus, each four-fermi term is replaced by a term
which is the product of the fermion bilinear and the auxiliary field, plus
a quadratic term in the auxiliary field. With judiciously chosen constants,
the original Lagrangian is regained by solving the equation of motion for
the auxiliary field and substituting this back into the Lagrangian.

In fact, our procedure will be a slight variant on this. Recall that our
goal is to show that a $\BF$ term arises in this model. Thus, it is clearly
necessary to introduce the auxiliary $B_\mn$ field. On the other hand, the
auxiliary $a_\mu$ field does not really interest us and we can simplify our
analysis by studying the theory using standard perturbation
theory techniques with respect to $g_1$, the coefficient of the Thirring
term. We will comment briefly below on the symmetric approach where
both four-fermi terms are eliminated in favour of auxiliary fields.

We are thus led to consider the following Lagrangian:
$$
\LL=-{1\over4e^2}\fmn^2-{1\over 2g_2}B_\mn^2 +\bar\psi (i\Dsl-m)\psi
+\bar\psi i\sigma^\mn\gamma_5\psi B_\mn
-{g_1\over2}(\bar\psi\gamma^\mu\psi)^2
\eqn\hi
$$
Although this model is of academic interest in particle physics since the
dipole moments of the fundamental particles are known to be extremely
small, it can be relevant to condensed matter systems. The quasi-particles
in anti-ferromagnetic, magneto-optic materials, for example, have large
dipole moments [\odell]. Massive U(1)-invariant fermionic field theories
commonly appear as effective low-energy models of frustrated
antiferromagnetic spin systems [\wenzee,\zee].
Furthermore, tensor order parameters,
although not common, do occur in physical systems such as $^3$He-A
[\schakel].

We are interested in the effective action obtained by integrating out the
fermions; thus we consider the following path integral:
$$
Z=\int[\DD\bar\psi][\DD\psi][\DD A_\mu][\DD B_\mn]
e^{i\int d^4x\LL(A,B,\bar\psi,\psi)}.
\eqn\eight
$$
The effective action $\geff[A,B]$ is obtained in principle by performing
the fermionic functional integral:
$$
Z=\int[\DD A][\DD B] e^{i\geff[A,B]},
\eqn\yes
$$
where
$$
e^{i\geff[A,B]}=e^{i\int d^4x
\left(-{1\over4e^2}\fmn^2-{1\over2 g_2}\bmn^2\right)}
e^{i\geff^{(f)}[A,B]}.
\eqn\no
$$
Here we have defined $\geff^{(f)}$ to be
the fermionic contribution to the effective action:
$$
e^{i\geff^{(f)}[A,B]}=\int[\DD\bar\psi][\DD\psi]
e^{i\int d^4x\bar\psi(i\parsl-m-\Asl+i\Bsl)\psi}
e^{-i{g_1\over2}\int d^4x(\bar\psi\gamma^\mu\psi)^2},
\eqn\maybe
$$
where $\Bsl\equiv\sigma^\mn\gamma_5 B_\mn$. Although the last term renders
the functional integral intractable, we can expand it as a power series in
$g_1$:
$$
e^{i\geff^{(f)}[A,B]}=\int[\DD\bar\psi][\DD\psi]
e^{i\int d^4x\bar\psi(i\parsl-m-\Asl+i\Bsl)\psi}
\left(1-i{g_1\over2}\int d^4x(\bar\psi\gamma^\mu\psi)^2+o(g_1^2)\right).
\eqn\maybee
$$
Defining $S_0$ as the part of the action bilinear in the fermion (\ie the
term independent of $g_1$), the term
linear in $g_1$ can be written
$$
\eqalign{
	i {g_1\over2}\int d^4x\int[\DD\bar\psi][\DD\psi]e^{i S_0}&
	\left(\bar\psi(x)\gamma^\mu\psi(x)\right)^2\cr
	&=\left(\int[\DD\bar\psi][\DD\psi]e^{i S_0}\right)
	i {g_1\over2}\int d^4x
	\langle0|\left(\bar\psi(x)\gamma^\mu\psi(x)\right)^2|0\rangle\cr}
\eqn\aaa
$$
where $|0\rangle$ is the ground state
of the action $S_0$. Recombining this with
the $(g_1)^0$ term,
$$
\eqalign{
	e^{i\geff^{(f)}}&=\left(1-i {g_1\over2}\int d^4x
	\langle0|\left(\bar\psi(x)\gamma^\mu\psi(x)\right)^2|0\rangle
	+\dots\right)\int[\DD\bar\psi][\DD\psi]e^{i S_0}\cr
	&\simeq e^{-i {g_1\over2}\int d^4x
	\langle0|\left(\bar\psi(x)\gamma^\mu\psi(x)\right)^2|0\rangle}
	\int[\DD\bar\psi][\DD\psi]e^{i S_0}.\cr}
\eqn\bbb
$$
So, to order $(g_1)^0$, $\geff^{(f)}$ is obtained via the functional
integral of the quadratic action $S_0$, while the correction to order $g_1$
is expressed in terms of the expectation value in the vacuum of $S_0$ of
the square of the current.

Let us first compute the $(g_1)^0$ contribution, which we call
${\geff^{(f)}}_0$. This is
$$
{\geff^{(f)}}_0=-i\Tr\log\left(i\parsl-m-\Asl+i\Bsl\right).
\eqn\ccc
$$
This can be decomposed
by separating off the
trace-log of the operator $i\parsl-m$ and expanding the resulting
logarithm. Thus, up to a field-independent (and therefore irrelevant),
infinite term,
$$
\eqalign{
	{\geff ^{(f)}}_0&=-i\Tr\log
	\left(1+{1\over i\parsl-m}(-\Asl+i\Bsl)\right)\cr
	&=\sum_{n=1}^\infty{i\over n}(-1)^n
	\Tr\left({1\over i\parsl-m}(-\Asl+i\Bsl)\right)^n,\cr}
\eqn\dix
$$
where the trace is over spinor indices
as well as momenta.

The first nonvanishing contribution to $\geff $ can be obtained
from evaluation of the $n=2$ term, which will be quadratic in the fields
$A$ and $B$. This term is
$$
{i\over2}\tr\int{d^4p\over(2\pi)^4}\bra{p}
{1\over i\parsl-m}(-\Asl+i\Bsl){1\over i\parsl-m}(-\Asl+i\Bsl)
\ket{p},
\eqn\onze
$$
where now the trace is only over spinor indices.

To evaluate this term, we need to separate the $x$-dependent and $p$-dependent
parts of the integrand, in order to obtain independent $x$ and $p$ integrals.
While this cannot be done exactly, a gradient expansion can be obtained with a
minimum of pain [\aitfra].
Keeping only terms to two derivatives, we get
three quadratic terms in the effective action:
$$
\eqalign{
	{\geff^{(f)}}_0[A,A]&=-{1\over48\pi^2}
	\log(\Lambda^2/m^2)
	\int d^4x\,\fmn^2,\cr
	{\geff^{(f)}}_0[A,B]&=-{m\over8\pi^2}
	\log(\Lambda^2/m^2)
	\int d^4x\,\eps^\mnrs\bmn F_\rs,\cr
	{\geff^{(f)}}_0[B,B]&={1\over 4\pi^2}\log(\Lambda^2/m^2)
	\int d^4x\left(m^2 \bmn^2+\bMN(i\partial)^2\bmn
	-4B^{\mu\sigma}i\partial_\mu i\partial_\nu {B^\nu}_\sigma\right).
	\cr}
\eqn\stuff
$$
Here we have assumed $\Lambda\gg m$, and have therefore
neglected terms which are finite, since they are dominated by the log.
At a more profound level, we have ensured gauge invariance with
respect to $A_\mu$ by Pauli-Villars regularization.

{}From \stuff, we see that we do indeed generate a $\BF$ term.
However, the third member of \stuff\ indicates
that there is a kinetic term for $B_\mn$
which is not of the desired form, \hey.
This implies that $\Gamma(B,B)$ is not invariant
under the generalized gauge invariance
$\bmn\to\bmn+\partial_\mu\Lambda_\nu-\partial_\nu\Lambda_\mu$ of the free
Lagrangian \un. This is, in fact, no surprise,
since the initial action \hi\ was not
invariant under such a transformation. Although expected, this
non-invariance is nonetheless alarming, since it is then not clear that the
argument for photon mass generation given in [\mass] still applies.
We will see below that under certain
circumstances, the offending gauge-non-invariant term in \stuff\ can be
neglected, in which circumstances the effective Lagrangian has the desired
interpretation, namely, the photon acquires a mass.

It remains to calculate the correction to order $g_1$, ${\geff^{(f)}}_1$,
which can be inferred from \bbb. We must compute the following
vacuum expectation value:
$$
\langle0|\left(\bar\psi(x)\gamma^\mu\psi(x)\right)^2|0\rangle.
\eqn\fff
$$
We use an approximation known as vacuum dominance to write
\fff\ in the following form:
$$
\left(\langle0|\bar\psi(x)\gamma^\mu\psi(x)|0\rangle\right)^2,
\eqn\ggg
$$
in which form it is readily computable as the square of the expectation
value of the current:
$$
\langle j^\mu(x)\rangle=\langle0|\bar\psi(x)\gamma^\mu\psi(x)|0\rangle
={\delta{\geff^{(f)}}_0 \over\delta A_\mu(x)}
={m\over4\pi^2}\log{\Lambda^2\over m^2}\eps^\mnrs\partial_\nu B_\rs(x),
\eqn\martin
$$
where an integration by parts has been performed. Thus
$$
\langle j^\mu(x)\rangle^2=-{m^2\over24\pi^4}\log^2{\Lambda^2\over m^2}
\hmnr\hMNR.
\eqn\ddd
$$
We are now in a position to reassemble all the terms in the effective
action which are bilinear in the fields and of two or fewer derivatives:
the tree level terms (the first of \no), the fermion contributions of order
$(g_1)^0$, \stuff, and the fermion contribution of order $(g_1)^1$,
obtained from \ddd. The result is
$$
\eqalign{
	\geff&=\int d^4x
	\bigg\{{1\over2g_2}\bmn^2\left({g_2m^2\over2\pi^2}
	\log{\Lambda^2\over m^2}-1\right)
	-{1\over4e^2}\fmn^2\left(1+{e^2\over12\pi^2}
	\log{\Lambda^2\over m^2}\right)\cr
	&\qquad-{m\over8\pi^2}
	\log{\Lambda^2\over m^2}\eps^\mnrs\bmn F_\rs
	+{g_1\over4}{m^2\over\pi^4}
	\log^2{\Lambda^2\over m^2}{1\over12}\hmnr\hMNR\cr
	&\qquad-{1\over4\pi^2}\log{\Lambda^2\over m^2}
	B^{\mu\sigma}\left(\partial^2g_\mn
	-4\partial_\mu\partial_\nu\right){B^\nu}_\sigma.\bigg\}.\cr}
\eqn\final
$$
This can be cleaned up somewhat with the following field redefinition
$$
\tilde\bmn={\sqrt{g_1}\over2}{m\over\pi^2}
\log{\Lambda^2\over m^2}\bmn;
\eqn\rashmi
$$
furthermore, we can drop the radiative correction to the photon kinetic
term to lowest order. We therefore arrive at
the following expression for the effective action:
$$
\eqalign{
	\geff[A,B] &=\int d^4x\biggl(-{1\over4e^2}\fmn^2
	+{1\over12}\tilde\hmnr\tilde\hMNR
	-{1\over4\sqrt{g_1}}\eps^\mnrs\tilde\bmn F_\rs\cr
	&\qquad+{2\tilde\bmn^2\over g_1g_2}
	\left({\pi^2\over m\log(\Lambda^2/m^2)}\right)^2
	\left({g_2m^2\over2\pi^2}\log{\Lambda^2\over m^2}-1\right)\cr
	&\qquad	-{\pi^2\over g_1m^2\log(\Lambda^2/m^2)}
	\tilde B^{\mu\sigma}\left(\partial^2 g_\mn
	-4\partial_\mu\partial_\nu\right){\tilde B^\nu}_\sigma\biggr)\cr}
\eqn\morr
$$
The first three terms are those shown in [\mass] to result in mass
generation for the photon. We must therefore
contend with the last two terms.

Ordinarily, the fourth term would be interpreted as a mass term for the
antisymmetric tensor field. However,
suppose we tune the cutoff in such a way that
the coefficient of this term vanishes.
Then the following gap equation is satisfied:
$$
m^2=\Lambda^2e^{-2\pi^2/m^2g_2}.
\eqn\gap
$$
This is a consistent choice since we have
assumed that the coupling $g_2$ is small, so that
the product of the exponential
with $\Lambda^2$ can indeed be of order $m^2$. A large value for $\Lambda^2$
implies weak coupling, so perturbation theory is valid.

We must also contend with the last term of \morr, which destroys the
generalized gauge invariance of the first three terms. The coefficient of
the last term, using \gap, is
$$
{\pi^2\over g_1m^2\log(\Lambda^2/m^2)}={g_2\over 2 g_1}
\eqn\ohno
$$
Assuming this ratio
is small allows us to drop this term relative to the kinetic
term for $\bmn$, resulting in the following effective action:
$$
\geff =\int d^4x\left\{-{1\over4e^2}\fmn^2
+{1\over12}\tilde\hmnr\tilde\hMNR
-{\eps^\mnrs\tilde\bmn F_\rs\over4\sqrt{g_1}}\right\}.
\eqn\lasst
$$
As shown in [\mass], this action
leads to a massive spin-1 vector field with
$$
M^2={1\over g_1}.
\eqn\jack
$$
We have thus accomplished our goal of demonstrating the existence of a
theory where this mass generation mechanism arises through radiative
corrections.

We note that the Lagrangian \lasst\ is invariant under both local U(1) gauge
transformations and under the generalized gauge transformations. It is
interesting to notice that we have restored generalized gauge invariance by
looking at the effective low-energy theory. Furthermore, the U(1) gauge
invariance is maintained for the massive photon by requiring that the extra
longitudinal mode transforms also under the electromagnetic gauge
invariance [\mass].

As we remarked above, the procedure we have used treated the two four-fermi
terms on different footing: the dipole-dipole term was eliminated by
introducing the auxiliary field $\bmn$, while the current-current term was
treated using conventional perturbation theory. This was done for reasons
of calculational simplicity, but it is perhaps worth describing very
briefly the
alternative, symmetric approach. One can imagine introducing a second
auxiliary field, $a_\mu$, which replaces the current-current term in the
same fashion; rather than \hi, we start from
$$
\LL=
-{1\over4e^2}\fmn^2-{1\over 2g_2}B_\mn^2-{1\over2g_1}a_\mu^2
+\bar\psi (i\Dsl-m)\psi
+\bar\psi i\sigma^\mn\gamma_5\psi B_\mn
+\bar\psi i\gamma^\mu\psi a_\mu.
\eqn\david
$$
If we integrate over  the fermion, using Pauli-Villars
regularization for the terms in the effective action involving
$A_\mu$, $a_\mu$ to keep the result gauge invariant with respect to
$A_\mu$,
we
obtain an effective action for the
fields $A_\mu$, $a_\mu$ and $\bmn$. There are now two possible procedures.
First, we could integrate over the auxiliary vector field $a_\mu$. Keeping
only two-derivative terms, one arrives eventually at the same result as
above, \final. Alternatively, we could integrate over $\bmn$, obtaining an
effective theory for $A_\mu$ and $a_\mu$. It turns out that if the same gap
equation \gap\ is satisfied, then the field strength derived from the
potential $A_\mu-ia_\mu$ is zero. We see then that the two potentials
differ only by a gauge transformation and again conclude that the
photon acquires a mass $M^2=1/g_1$ via the Stuckelberg compensating field
argument. Thus, the non-symmetric way we treated
the two four-fermi terms in the original Lagrangian \six\ is equivalent to
the somewhat more involved, but more natural-appearing, symmetric approach.

As a final note, it is perhaps more interesting in terms of particle
physics phenomenology to discuss the non-abelian case, since massive
non-abelian gauge fields are ubiquitous
in that context. Indeed, starting with a nonabelian
version of \hi,
$$
\LL={1\over4e^2}\Tr\fmn^2+\bar\psi(i\Dsl-m)\psi
+{1\over 2g_2}\Tr(\bmn^iT^i)^2
-{g_1\over2}(\bar\psi\gamma_\mu{\bf 1}\psi)^2
+i(\bar\psi\sigma^\mu\gamma_5\bmn^iT^i)\psi,
\eqn\nonab
$$
where now $D_\mu=\partial_\mu+iA_\mu^iT^i$,
we can integrate over the fermion to obtain an effective action for the
non-abelian gauge field $A_\mu^i$ and $B_\mn^i$.
After some work, one finds that the effective action to two derivatives is
of a similar structure to the abelian case: one has kinetic terms for
the gauge field and for $B_\mn^i$, a term quadratic in $B$
with no derivatives
(the coefficient of which is set to zero by tuning the cutoff which
arises), and a term of the form $\BF$. This is exactly the situation
analysed in [\namass], wherein it was concluded that this combination of
terms results in a mass for the non-abelian gauge bosons. We
conclude, therefore, that such four-fermi terms can indeed generate masses
for non-abelian gauge fields as well.

We gratefully acknowledge useful
discussions with V.P. Nair.
This work was supported in part by the Natural Science and
Engineering Research Council of Canada, the Fonds F.C.A.R. du Qu\'ebec,
and the U.S. National Science Foundation.

\refout

\end